\documentclass[%
 reprint,
superscriptaddress,
 amsmath,amssymb,
 aps,
]{revtex4-2}
\usepackage{graphicx} 
\usepackage{tikz}
\usepackage{natbib}%
\usepackage{hyperref}
\usepackage{subfigure}
\usepackage{amsmath}
\usepackage{comment}

\usepackage{svg}
\usepackage{booktabs}
\usepackage{multirow}
\usepackage{array}
\usepackage{makecell}
\setcitestyle{numbers,super,square}

\begin{document}

\title{Temperature-dependent Optical and Polaritonic Properties of hBN-encapsulated Monolayer TMDs}

\author{Matan Meshulam}
\affiliation{School of Physics, Faculty of Exact Sciences, Tel Aviv University, Tel Aviv 6997801, Israel}
\affiliation{School of Electrical Engineering, Faculty of Engineering, Tel Aviv University, Tel Aviv 6997801, Israel}
\author{Anabel Atash Kahlon}
\affiliation{School of Electrical Engineering, Faculty of Engineering, Tel Aviv University, Tel Aviv 6997801, Israel}
\author{Yonatan Gershuni}
\affiliation{School of Physics, Faculty of Exact Sciences, Tel Aviv University, Tel Aviv 6997801, Israel}
\affiliation{School of Electrical Engineering, Faculty of Engineering, Tel Aviv University, Tel Aviv 6997801, Israel}
\author{Thomas Poirier}
\affiliation{Tim Taylor Department of Chemical Engineering, Kansas State University, Manhattan, KS 66506, USA}
\author{Thomas Poirier}
\affiliation{Tim Taylor Department of Chemical Engineering, Kansas State University, Manhattan, KS 66506, USA}
\author{Seth Ariel Tongay}
\affiliation{Materials Science and Engineering, School for Engineering of Matter,
Transport and Energy, Arizona State University, Tempe, Arizona 85287, USA}
\author{Itai Epstein}
\email{itaieps@tauex.tau.ac.il}
\affiliation{School of Electrical Engineering, Faculty of Engineering, Tel Aviv University, Tel Aviv 6997801, Israel}
\affiliation{Center for Light-Matter Interaction, Tel Aviv University, Tel Aviv 6997801, Israel}
\affiliation{QuanTAU, Quantum Science and Technology Center, Tel Aviv University, Tel Aviv 6997801, Israel}


\begin{abstract}
Monolayer transition metal dichalcogenides (TMDs) support robust excitons in the visible to near-infrared spectral range. Their reduced dielectric screening results in large binding energies, and combined with a direct bandgap in monolayer form, these excitons dominate the optical response of TMDs. In this work, we present a comprehensive investigation of the temperature-dependent optical and polaritonic properties of high-quality, flux-grown $\mathrm{WS_2}$, $\mathrm{MoS_2}$, $\mathrm{WSe_2}$, and $\mathrm{MoSe_2}$  monolayers, encapsulated by hBN, in the full temperature range between $5-300 \mathrm{K}$. Using reflection spectroscopy measurements, we evaluate and compare the optical and polaritonic constituents of the TMD excitons in terms of oscillator strength, linewidth and negative permittivity. We find that it is $\mathrm{MoSe_2}$ that exhibits the most pronounced optical and polaritonic response, stemming from its rapid linewidth narrowing at low temperatures, as compared to the temperature-dependent response of the other TMDs. In addition, we find that all four TMDs exhibits a temperature-dependent negative real part permittivity, thus supporting surface-exciton-polaritons. We derive their dispersion relation, confinement factors and losses, similarly revealing that $\mathrm{MoSe_2}$ exhibit enhanced polaritonic properties. These findings establish a comparative framework for understanding the optical and polaritonic properties of monolayer TMDs, with implications on their utilization in optoelectronic devices based on 2D semiconductors.
\end{abstract}

\maketitle

\section{Introduction}\label{sec1}

\begin{figure*}
    \centering
    \includegraphics[width=0.85\textwidth]{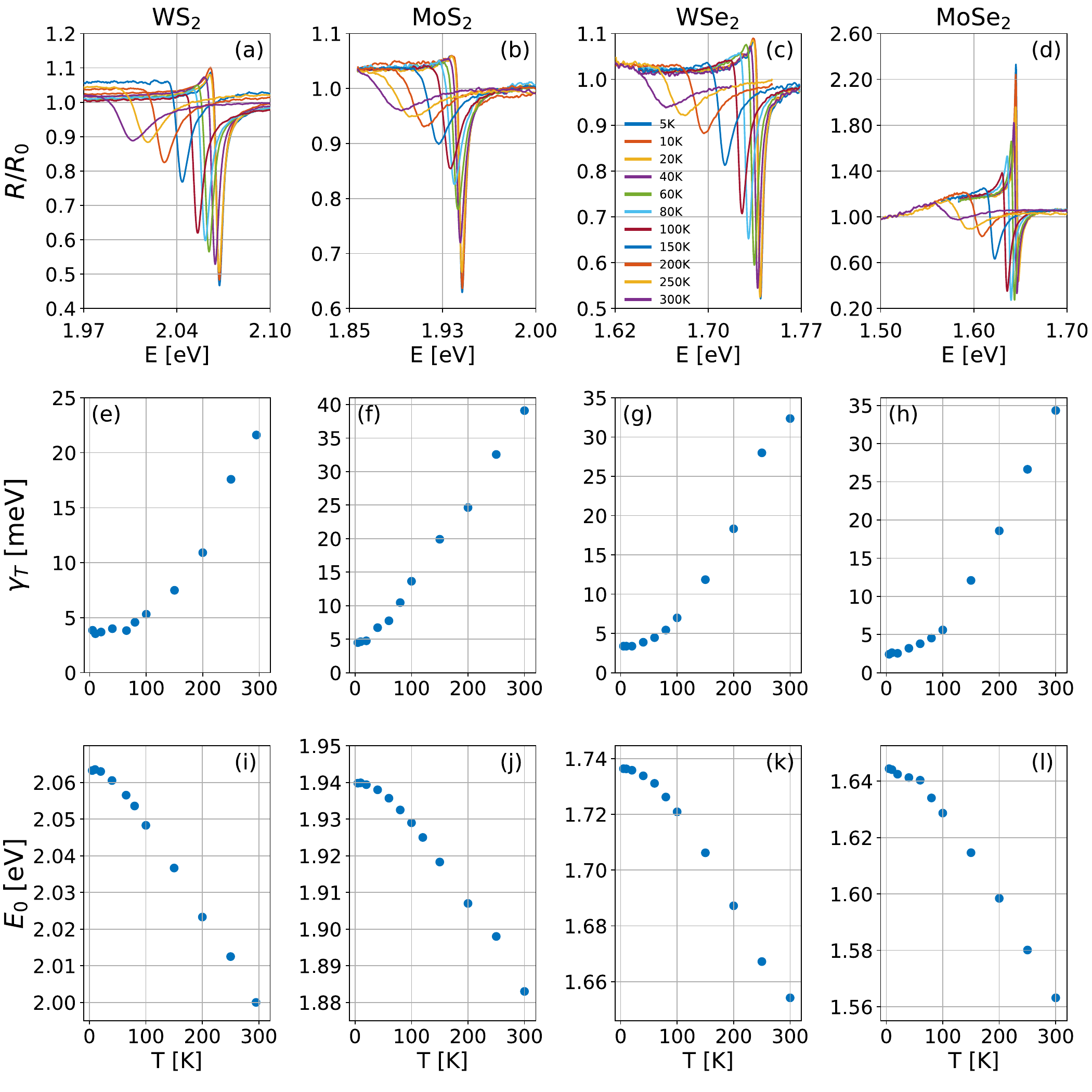}
    \caption{Temperature-dependent optical properties of hBN-encapsulated monolayer TMDs ($\mathrm{WS_2}$, $\mathrm{MoS_2}$, $\mathrm{WSe_2}$, $\mathrm{MoSe_2}$) from $5\mathrm{K}$ to $300\mathrm{K}$. (a-d) Reflection contrast spectra showing the evolution of excitonic resonances with temperature. (e-h) Exciton linewidth evolution with temperature. (i-l) Exciton energy positions as a function of temperature.}
    \label{fig: Fig 1}
\end{figure*}

Monolayer transition metal dichalcogenides (TMDs) have attracted considerable attention in recent years due to their unique excitonic properties, such as strong spin-orbit coupling \cite{PhysRevLett.105.136805, Xiao2012_SpinValley_MoS2}, valley-specific optical selection rules \cite{Mak_He_Shan_Heinz_2012, PhysRevB.86.081301}, and large binding energies due to reduced screening \cite{Chernikov2014_BindingEnergies, Liu2014_Ellipsometry_MonolayerTMDs, Kravets2017_Ellipsometry_TMDs, RevModPhys.90.021001}. Together with the direct bandgap nature in monolayer form, these lead to the fact that excitons in TMDs dominate their optical response \cite{Liu2014_Ellipsometry_MonolayerTMDs, RevModPhys.90.021001}. \par

The excitonic linewidth plays a defining role in the optical properties of monolayer TMDs, being affected by both intrinsic and extrinsic influences, which contribute to its broadening via homogeneous and inhomogeneous broadening \cite{Moody2015_ExcitonDynamics, Cadiz2017_PRX_Linewidth,Selig2016_PhononAssistedLinewidth}. While homogeneous broadening arises from intrinsic effects, inhomogeneous broadening results from extrinsic perturbations, such as local strain, charge traps, dielectric variations, etc. These influences depend on the TMD’s quality and environment, and can significantly affect the linewidth \cite{PhysRevB.93.205423, Moody:16, Selig2016_PhononAssistedLinewidth, Liu2016-jd, Raja2017-dl, Ajayi2017_IntrinsicPL_Linewidth, PhysRevB.98.115308, Rhodes2019_Disorder_vdW, Shree_2020}. Such effects are particularly prominent in commercially available materials, where large defect concentrations lead to increased scattering and low optical quality. In contrast, high-quality flux-grown TMDs contain fewer impurities, resulting in narrower linewidths \cite{Ajayi2017_IntrinsicPL_Linewidth, Rhodes2019_Disorder_vdW, PhysRevApplied.14.021002}. To further reduce broadening effects, monolayer TMDs are commonly encapsulated in hexagonal-boron-nitride (hBN), which reduces environmental effects, and together with the lowering of temperature, which reduces exciton-phonon interaction, allows for approaching the homogeneous exciton linewidth \cite{Moody2015_ExcitonDynamics,  PhysRevLett.116.127402, Cadiz2017_PRX_Linewidth, Cadiz2017_PRX_Linewidth, Ajayi2017_IntrinsicPL_Linewidth, Epstein2020-ae, AtashKahlon2025_PureDephasing}.  \par

In such high-quality flux-grown, hBN-encapsulated TMD heterostructures, this linewidth narrowing can lead to the emergence of a negative real part in the in-plane permittivity, giving rise to polaritonic effects, such as surface polaritons based on excitonic response \cite{PhysRevB.94.195418, Karanikolas:19, PhysRevResearch.2.033141, Epstein2020_InPlaneEP, Li2021-mb, Ajayi2017_IntrinsicPL_Linewidth, Gershuni2024_ExcitonVsPlasmon, Eini2025_GrapheneExcitonPRL}. Moreover, since the out-of-plane permittivity remains positive under these conditions, the material can exhibit a hyperbolic repose, and the emergence of hyperbolic-exciton-polaritons \cite{Fang2005-sv, Pacifici2007-bs, Anker2008-js, Kats2025_HyperbolicSuperlattice}. Specifically in TMDs, coupling of surface polaritons to the valley degree-of-freedom has also been predicted, enabling valley-selective and directional propagation \cite{PhysRevB.106.L201405, Gershuni2024_ExcitonVsPlasmon}. Such surface and hyperbolic polaritons are known for their ability to confine electromagnetic fields to sub-wavelength scales, well below the diffraction limit, and enhance the local electromagnetic fields by orders of magnitude. These confinement factors stem from the large momentum carried by such surface-polaritons, which directly translates to a reduction of the polariton's wavelengths. Consequently, polaritons in these systems facilitate remarkably strong optical field intensities, making them highly important for light-matter interactions at the nano-scale \cite{Nikolajsen2004-of, Fang2005-sv, Pacifici2007-bs, Anker2008-js, Fu2012-uv, Wang2013-hu, Poddubny2013-iz, Shekhar2014-tw, FERRARI20151, Hu2020-yg, Wang2024-qt, Ma2024-qe}.\par 

While the constituents of the optical response of excitons in monolayer TMDs have been studied, it has been focused mostly on commercially-grown materials of large defect density, and/or non-encapsulated TMDs, and/or at room temperature \cite{Liu2014_Ellipsometry_MonolayerTMDs, PhysRevB.90.205422, Kravets2017_Ellipsometry_TMDs, Liu2020_TMD_OpticalConstants_Tdependence, Munkhbat2022_MultilayerTMD_OpticalConstants, Nguyen2024_WSe2_Dielectric_Tdependence, nano8090725}, where the optical response is of a relatively poor nature. Thus, the exploration of the optical and polaritonic constituents of high-quality monolayer TMD heterostructures has not yet been conducted. In this work, we present a comprehensive investigation of the temperature-dependent optical and polaritonic properties of high-quality, flux-grown $\mathrm{WS_2}$, $\mathrm{MoS_2}$, $\mathrm{WSe_2}$, and $\mathrm{MoSe_2}$ monolayers, encapsulated by hBN, in the full temperature range between $5-300 \mathrm{K}$. Using reflection spectroscopy measurements, we evaluate and compare the optical and polaritonic properties of each TMD excitons in terms of oscillator strength, linewidth and negative permittivity. We find that it is $\mathrm{MoSe_2}$ that exhibits the most pronounced properties, stemming from its rapid linewidth narrowing at low temperatures, as compared to the temperature-dependent response of the other TMDs. These findings are especially important owing to the common view that it is $\mathrm{WS_2}$ that exhibits the strongest optical response and largest oscillator strength. We explain this observation as being previously obscured in other studies focused on either room-temperature analysis, and/or on non-encapsulated monolayers, and/or low-quality TMD material. In addition, we find that all four TMDs exhibits a temperature-dependent negative real part of their permittivities, and can thus exhibit polaritonic response. As an exemplary case, we study the supported surface-exciton-polaritons in the heterostructures and derive their dispersion relation, confinement factors and losses, taking into account nonlocal corrections to the excitonic response. Similarly, we find that $\mathrm{MoSe_2}$ exhibits enhanced polaritonic properties, as compared to the other TMDs, owing to its narrow linewidth and accompanying large oscillator strength.

\section{Results}\label{sec2}
To investigate the optical and polaritonic properties of the four monolayer TMD heterostructures, we measure their temperature-dependent reflection and photoluminescence (PL) spectra. From the reflection data we extract the complex frequency-dependent permittivity for each temperature, which characterizes its optical response and its constituents. The extracted permittivity is then used to evaluate and compare each material's optical and polaritonic response and derive the corresponding dispersion relation.

Figure \ref{fig: Fig 1} presents the measured temperature-dependent reflection contrast, defined as $\frac{R}{R_0}$, where $R$ is the reflectance from the complete heterostructure and $R_0$ is the reflectance from the heterostructure without the monolayer TMD. It can be seen in Figures \ref{fig: Fig 1} (a-d) that in all four TMD heterostructures the excitonic resonance broadens and decreases in amplitude with increasing temperature, owing to exciton-phonon interaction, which agrees well with previous reports \cite{Wang2015-rq, Moody2015_ExcitonDynamics, Selig2016_PhononAssistedLinewidth, PhysRevLett.116.127402, Scuri2018-hk, Liu2020_TMD_OpticalConstants_Tdependence, AtashKahlon2025_PureDephasing}. Figures \ref{fig: Fig 1} (e-h) show the excitonic linewidth as a function of temperature, as extracted from the FWHM of PL measurements, reaching at $5\mathrm{K}$ close to their homogeneous linewidth, as reported in previous reports \cite{Cadiz2017_PRX_Linewidth, Ajayi2017_IntrinsicPL_Linewidth} in high-quality heterostructures. 

It can further be seen that the total linewidth, $\gamma_T$, increases linearly at low temperatures and more rapidly above approximately $100 \mathrm{K}$, reflecting the growing influence of exciton–phonon interactions. This behavior is well described by the phenomenological relation: $\gamma_T(T) = \gamma(0) + c_1 T + \frac{c_2}{e^{{\Omega}/{k_B T}} - 1}$, consistent with previous reports \cite{Selig2016_PhononAssistedLinewidth, PhysRevLett.116.127402, Cadiz2017_PRX_Linewidth,  AtashKahlon2025_PureDephasing}, where $\gamma(0)$ is the zero-temperature linewidth in the absence of phonons, $c_1$ and $c_2$ coupling coefficients to acoustic and optical phonons, respectively, and $\Omega$ denotes the average optical phonon energy. For completeness, we show in figures \ref{fig: Fig 1} (i-l) the temperature evolution of the resonances' energies, exhibiting the expected shift in the exciton energy with temperature, coinciding with previous reports \cite{PhysRevLett.116.127402, Epstein2020-ae}.

From the extracted linewidths in Figures \ref{fig: Fig 1} (e-h), it is seen that at $300\mathrm{K}$ $\mathrm{WS_2}$ exhibits a notably narrower linewidth compared to the other TMDs, implying why it has been previously identified as possessing the most pronounced optical properties \cite{PhysRevB.90.205422, nano8090725, Liu2014_Ellipsometry_MonolayerTMDs, Liu2020_TMD_OpticalConstants_Tdependence, Munkhbat2022_MultilayerTMD_OpticalConstants, Kravets2017_Ellipsometry_TMDs}. As the temperature decreases, it can be seen that the linewidths of all TMDs gradually narrow toward their respective minimal values. However, in comparison to the other TMDs, $\mathrm{MoSe_2}$ attains narrow linewidth values already at around $\sim100\mathrm{K}$ and continues narrowing with decreasing temperature, reaching almost half the linewidth of the other TMDs at $5\mathrm{K}$. 

This distinct behavior of $\mathrm{MoSe_2}$ directly influences the reflection amplitude of the excitonic resonance, associated with its oscillator strength. It can be seen in Figure \ref{fig: Fig 1}(d) that between $300-100\mathrm{K}$, the change in reflection amplitude of $\mathrm{MoSe_2}$ is similar to that of the other TMDs. However, around $\sim100\mathrm{K}$ and below, it drastically increases, reaching roughly three times larger amplitude at $5\mathrm{K}$ compared to the other TMDs. This behavior directly stems from the linewidth narrowing of $\mathrm{MoSe_2}$ at this temperature range, whereas the linewidths of the other TMDs remain around $\sim5\mathrm{meV}$ or above (Figures \ref{fig: Fig 1} (e-h)). This increased amplitude implies that $\mathrm{MoSe_2}$ possesses the strongest optical response among the investigated TMDs under these conditions, as will be further demonstrated in the following section.

To eliminate the effects of hBN thicknesses and the substrate response on the reflection spectra, next we extract the frequency-dependent permittivity of the monolayer TMDs, which characterizes its response to an electromagnetic field. For 2D excitons, this is commonly modeled by a Lorentzian complex susceptibility given by \cite{PhysRevB.90.205422, Epstein2020_InPlaneEP, Epstein2020-ae, Gershuni2024_ExcitonVsPlasmon}:

\begin{equation} \label{eq:1} \chi_{\perp}(\omega) = \chi_{bg} - \frac{c}{\omega_0 d_0}\frac{\gamma_r}{\omega - \omega_0 + i\frac{\gamma_{nr}}{2}}. \end{equation}

Here, $\chi_{bg}$ is the background susceptibility, accounting for contributions from higher-energy optical transitions, $c$ is the speed of light, $\omega_0$ is the exciton resonance frequency, $d_0$ is the monolayer thickness, $\gamma_r$ is the radiative decay rate, describing exciton decay via radiative recombination, and $\gamma_{nr}$ is the non-radiative decay rate, characterizing exciton population loss via non-radiative processes \cite{PhysRevB.90.205422,  PhysRevLett.118.113601, Scuri2018-hk, PhysRevLett.120.037401, AtashKahlon2025_PureDephasing}. \par

To extract $\gamma_r$ and $\gamma_{nr}$, we fit the reflection contrast of the heterostructure using a Transmission Line Model (TLM) simulations \cite{eini2025transmissionlinemodel2d, PhysRevB.106.L201405, AtashKahlon2025_PureDephasing, eini2025stronglycoupledexcitonhyperbolicphononpolaritonhybridized, Eini2025_GrapheneExcitonPRL, Kats2025_HyperbolicSuperlattice}, and obtain the complex permittivity via $\epsilon(\omega) = 1 + \chi_{\perp}(\omega)$.\par 

\begin{figure*}
    \centering
    \includegraphics[width=0.85\textwidth]{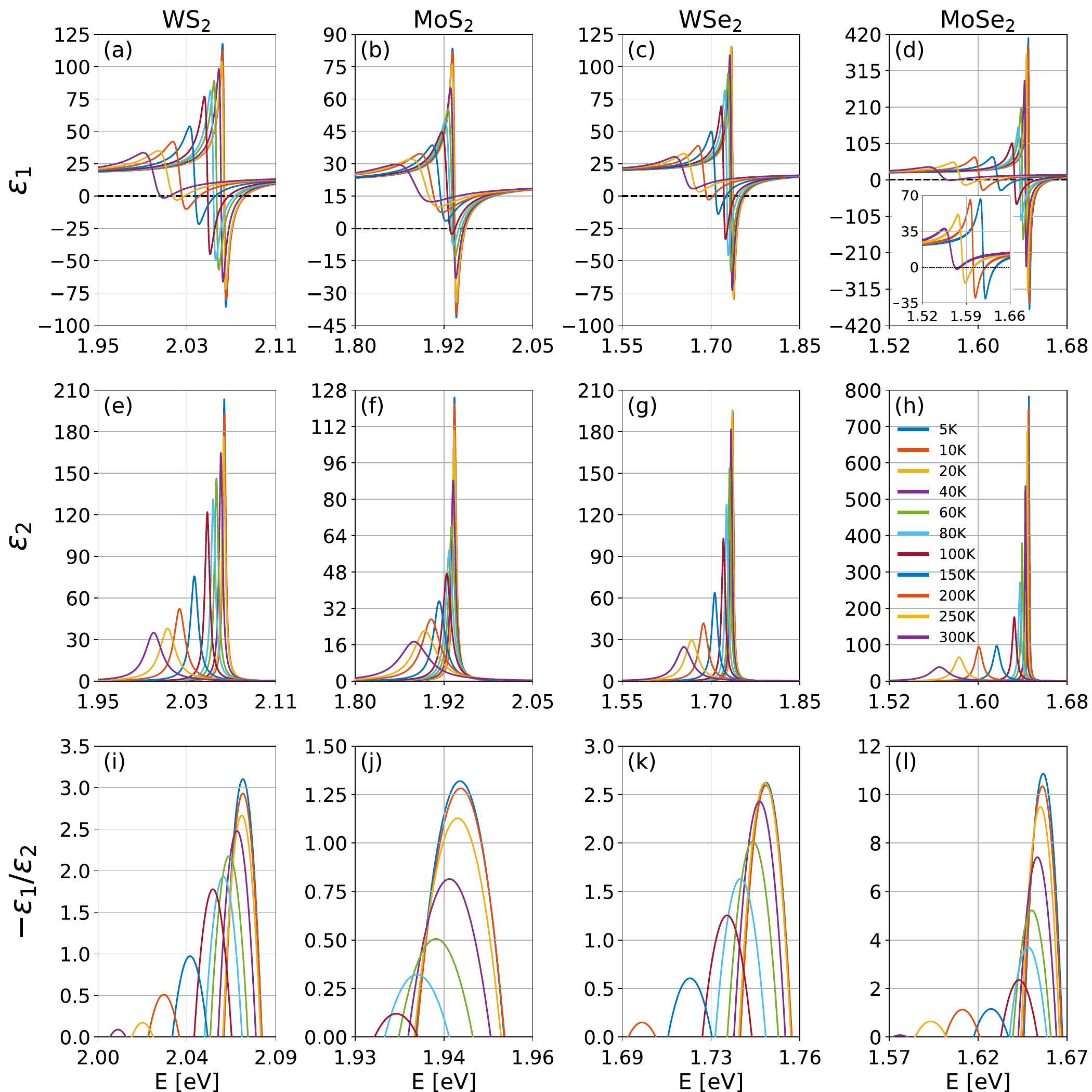}
    \caption{Complex permittivity of hBN-encapsulated monolayer TMDs at various temperatures. (a-d) Real part of permittivity ($\epsilon_1$), black dashed horizontal lines mark zero crossing. Inset shows a zoom-in on the energy range containing $300\mathrm{K} - 150\mathrm{K}$. (e-h) Imaginary part of the permittivity ($\epsilon_2$). (i-l) The ratio $-\frac{\epsilon_1}{\epsilon_2}$, indicating spectral regions with optimal conditions for polaritonic response.}
    \label{fig: Fig.2}
\end{figure*}

\begin{figure*}
    \centering
    \includegraphics[width=0.85\textwidth]{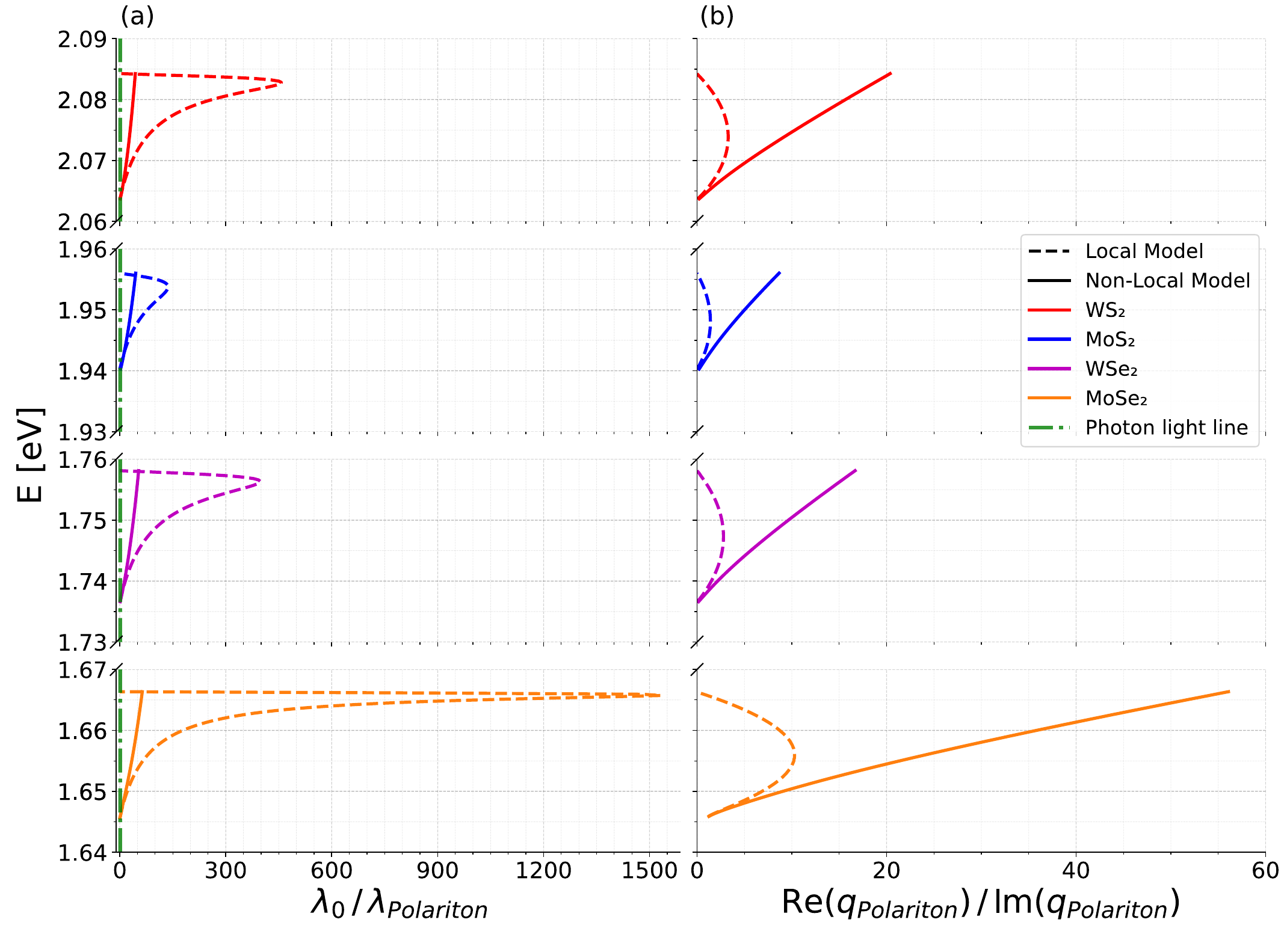}
    \caption{Polaritonic properties of surface-exciton-polaritons supported by the four TMDs. (a) Confinement factor, $\lambda_0/\lambda_{Polariton}$, under the local and non-local models in (dashed and full lines, respectively). Vertical dashed-dot line indicates confinement factor of unity, corresponding to the light-line. (b) The loss figure of merit, $Re(q_{Polariton})/Im(q_{Polariton})$}
    \label{fig: Fig.3}
\end{figure*}

Figure \ref{fig: Fig.2} presents the extracted permittivity of the four monolayer TMDs for all measured temperatures. It can be seen that the amplitude of the real part of the permittivity, $\epsilon_1$, increases with decreasing temperature (Figure \ref{fig: Fig.2} (a-d)), while the linewidth of the imaginary part of the permittivity, $\epsilon_2$, narrows with decreasing temperatures (Figure \ref{fig: Fig.2} (e-h)). This behavior can be understood from the Lorentzian nature of the excitonic resonance, upholding Kramers-Kronig relations.

It is also evident that the amplitude of $\epsilon_1$ differs significantly among the studied monolayer TMDs (Figures \ref{fig: Fig.2} (a-d)). Since the amplitude of $\epsilon_1$ is directly influenced by its linewidth, its behavior across different TMDs and temperatures closely follows those observed in the reflection contrast measurements discussed above. While all the TMDs exhibit similar amplitudes of $\epsilon_1$ at temperatures above $100\mathrm{K}$, below this threshold the amplitude in $\mathrm{MoSe_2}$ rapidly increases, resulting in values roughly over three times greater than those of the other TMDs (Figures \ref{fig: Fig.2} (a-d)). This increase corresponds directly to the pronounced linewidth narrowing of $\epsilon_2$ as compared to the other TMDs (Figures \ref{fig: Fig.2} (e-h)) at the same spectral range. We can thus conclude that $\mathrm{MoSe_2}$ exhibits the most pronounced optical response among the four TMDs under these conditions, and note that this observation contrasts earlier studies conducted either at room-temperature, and/or on lower-quality and/or non-encapsulated TMDs, which have generally identified $\mathrm{WS_2}$ as possessing the strongest optical response \cite{PhysRevB.90.205422, nano8090725, Liu2014_Ellipsometry_MonolayerTMDs, Liu2020_TMD_OpticalConstants_Tdependence, Munkhbat2022_MultilayerTMD_OpticalConstants, Kravets2017_Ellipsometry_TMDs}.

Figures \ref{fig: Fig.2} (e–h) present the temperature-dependent evolution of the imaginary part of the permittivity ($\epsilon_2$), which quantifies absorption losses in the material. As temperature decreases, the excitonic resonance narrows, and the amplitude of $\epsilon_2$ simultaneously increases. $\mathrm{MoSe_2}$ exhibits the highest peak values of $\epsilon_2$, consistent with its large oscillator strength, which also results in the highest resonance amplitude in $\epsilon_1$. This demonstrates that the observed relationship between increased optical response and absorption losses is consistent across all investigated materials.

The temperature dependency of the excitonic resonance and the associated change in the amplitude of $\epsilon_1$ also impacts its ability to acquire negative values, thereby influencing the polaritonic properties of the TMDs. In particular, whether $\epsilon_1$ can acquire negative values depends on the interplay between the background susceptibility (setting a positive baseline for $\epsilon_1$) and the amplitude of the excitonic resonance. When the linewidth is large, the excitonic resonance is broad and its amplitude small, insufficient to reduce $\epsilon_1$ below zero, as in the case of low-quality commercially-grown, and/or non-encapsulated TMDs, and/or at room temperature \cite{PhysRevB.90.205422, Munkhbat2022_MultilayerTMD_OpticalConstants, Liu2014_Ellipsometry_MonolayerTMDs}. In contrast, at narrow linewidths achieved by using our high-quality TMDs, hBN encapsulated at low enough temperatures, and even at room temperature (for $\mathrm{WS_2}$ and $\mathrm{MoSe_2}$), the excitonic contribution enables $\epsilon_1$ to acquire negative values.

Among the four TMDs, $\mathrm{MoSe_2}$ exhibits the largest amplitude of negative $\epsilon_1$ at $5\mathrm{K}$, up to four times that of other TMDs, followed by $\mathrm{WS_2}$ and $\mathrm{WSe_2}$ which display comparable results per temperature, and $\mathrm{MoS_2}$ which displays significantly less negative values in $\epsilon_1$ throughout the entire temperature range. Differences can also be observed in the onset temperature for negative permittivity. While $\mathrm{MoSe_2}$ and $\mathrm{WS_2}$ exhibit slightly negative values even at room temperature, $\mathrm{WSe_2}$ and $\mathrm{MoS_2}$ only begin to show negative $\epsilon_1$ below $200\mathrm{K}$ and $100\mathrm{K}$ accordingly. The emergence of negative real permittivity thus satisfies the key requirement for surface-polariton formation.\par

\begin{table*}
\centering
\small
\setlength{\tabcolsep}{6pt}
\renewcommand{\arraystretch}{1.25}
\begin{tabular}{|l|c|*{11}{c|}}
\hline
\multicolumn{2}{|c|}{} & \multicolumn{11}{c|}{\textbf{Temperature (K)}} \\
\hline
\multirow{2}{*}{WS$_2$} 
 & $\gamma_r$    & \multicolumn{11}{c|}{2.05} \\
\cline{2-13}
 & $\gamma_{nr}$ & 3.07 & 3.24 & 3.54 & 3.79 & 4.28 & 4.78 & 5.15 & 8.35 & 12.20 & 16.86 & 18.43 \\
\hline
\multirow{2}{*}{MoS$_2$} 
 & $\gamma_r$    & \multicolumn{11}{c|}{2.13} \\
\cline{2-13}
 & $\gamma_{nr}$ & 5.32 & 5.44 & 6.00 & 7.53 & 9.75 & 11.58 & 14.12 & 19.27 & 31.02 & 38.32 & 48.99 \\
\hline
\multirow{2}{*}{WSe$_2$} 
 & $\gamma_r$    & \multicolumn{11}{c|}{2.00} \\
\cline{2-13}
 & $\gamma_{nr}$ & 3.80 & 3.84 & 3.84 & 4.09 & 4.85 & 5.85 & 7.28 & 11.83 & 18.28 & 26.15 & 31.57 \\
\hline
\multirow{2}{*}{MoSe$_2$} 
 & $\gamma_r$    & \multicolumn{11}{c|}{1.94} \\
\cline{2-13}
 & $\gamma_{nr}$ & 1.00 & 1.05 & 1.14 & 1.45 & 2.03 & 2.81 & 4.30 & 7.84 & 8.00 & 11.50 & 20.00 \\
\hline
\end{tabular}
\caption{Summary of $\gamma_r$ and $\gamma_{nr}$ values at temperatures ranging from $5\mathrm{K}$ to $300\mathrm{K}$, extracted from the reflection contrast measurements via a TLM, and used to obtain the permittivity values presented in Figure \ref{fig: Fig.2}. The displayed $\gamma_r$ values are the average over all temperatures for each material. $d_0$ was taken to be $d_{0,\mathrm{WS_2}} = 6.18\,\mathring{\mathrm{A}}$, $d_{0,\mathrm{MoS_2}} = 6.15\,\mathring{\mathrm{A}}$, $d_{0,\mathrm{WSe_2}} = 6.49\,\mathring{\mathrm{A}}$, $d_{0,\mathrm{MoSe_2}} = 6.46\,\mathring{\mathrm{A}}$ \cite{PhysRevB.90.205422}}
\label{table: 1}
\end{table*}

Figures \ref{fig: Fig.2} (i-l) present the ratio $-\epsilon_1/\epsilon_2$, quantifying a figure of merit for the optimal polaritonic properties of the material. This ratio takes into account the tradeoff between the TMDs' ability to support surface-exciton-polaritons, which is quantified by the negativity of $\epsilon_1$, and losses due to absorption in the material, which dissipates the energy of the propagating mode, quantified by $\epsilon_2$ \cite{Epstein2020_InPlaneEP}. In all materials, the ratio peaks at energies slightly above the excitonic resonance, where $\epsilon_1$ remains negative and $\epsilon_2$ has decreased from its maximum. At $5 \mathrm{K}$, each material displays a narrow spectral window where the ratio remains positive and near its peak value. As temperature increases, this window broadens, and the ratio gradually decreases, reflecting the material-specific evolution of $\epsilon_1$ and $\epsilon_2$ and its corresponding polaritonic qualities. It can be seen that among the four TMDs, $\mathrm{MoSe_2}$ exhibits the highest peak value, indicating its high potential for polaritonic response. \par

For convenience, the extracted decay rates used to obtain the permittivity of the different TMDs using Eq. \ref{eq:1} across all measured temperatures are summarized in Table \ref{table: 1}.\par


Having established that monolayer TMDs can support surface-exciton-polaritons, we now analyze the polaritonic properties of these modes by numerically computing their dispersion relation in each material, following the methodology outlined in Gershuni et al.\cite{Gershuni2024_ExcitonVsPlasmon}. As surface polaritons commonly carry large momentum, it is required to take into account the dependency of the permittivity on momentum, rather than solely on frequency as in Eq. \ref{eq:1}. This non-local correction incorporates a momentum-dependent kinetic energy term, thus introducing a momentum-dependent term to the permittivity \cite{PhysRev.132.563, PhysRevB.10.1447, Eini2025_GrapheneExcitonPRL}:

\begin{equation} \label{eq:2} E_{ex} \approx \omega_0 \rightarrow \omega_0 + \frac{\hbar^2 k^2}{2M}, \end{equation}
where $M$ is the excitonic effective mass \cite{Yu2010-cz}, and $k$ is the excitonic momentum. This correction modifies the susceptibility of monolayer TMDs presented in \ref{eq:1} to the form:
\begin{equation} \label{eq:3} \chi_{\perp}(\omega) = \chi_{bg} - \frac{c}{\omega_0 d_0}\frac{\gamma_r}{\omega - \left(\omega_0 + \frac{\hbar^2 k^2}{2M}\right) + i\frac{\gamma_{nr}}{2}}. \end{equation}

Importantly, this correction explicitly depends on the excitonic effective mass, which varies across different TMD materials due to changes in the effective masses of electrons and holes \cite{Kormanyos_2015, Hong_Koshino_Senga_Pichler_Xu_Suenaga_2021}. Using the effective masses reported by Hong et al. via the q-EELS method  \cite{Hong_Koshino_Senga_Pichler_Xu_Suenaga_2021}, we quantify the polaritonic behavior by extracting two material-specific polaritonic properties: the confinement factor and the propagation loss figure of merit. The confinement factor, defined as the ratio $\lambda_0/\lambda_{Polariton}$, where $\lambda_0$ is the free-space photon wavelength, and $\lambda_{Polariton}$ is the wavelength of the surface-exciton-polariton, quantifies the degree to which the polaritonic modes spatially confine electromagnetic fields compared to their free-space wavelength. The propagation losses, defined as $Re(q_{Polariton})/Im(q_{Polariton})$, where $q_{polariton}$ is the polaritonic momentum, intuitively translates to the number of wavelengths the mode propagates before decaying.

Figure \ref{fig: Fig.3} displays the confinement factors and propagation losses of the surface-exciton-polaritons supported by the four TMDs at $5\mathrm{K}$, as the polaritonic response is strongest at this temperature for all TMDs (Figures \ref{fig: Fig.2} (a-d)). Under the local model, the confinement factor varies significantly among the different TMDs, reaching values exceeding two orders of magnitude, implying that the polaritonic wavelength is comparable to the excitonic Bohr radius, which is not plausible. When non-local corrections are included, the confinement factor of all TMDs yield similar values. This behavior arises from the varying influence of the non-local corrections between TMDs, due to the differences in their excitonic masses used in Eq. \ref{eq:2}. This mass dependence also leads to the ordering of TMDs based on maximal confinement factors to differ from their ordering based on oscillator strengths, inferred from the amplitudes of $\epsilon{1,2}$ (Figures 2 (a-d),(e-h)).\par

Figure \ref{fig: Fig.3} (b) illustrates and compares the propagation losses across the different TMDs, exhibiting that employing the non-local correction provides significantly longer propagation lengths, stemming from the decrease in the confinement factors. In particular, $\mathrm{MoSe_2}$ demonstrates propagation loss exceeding $50$, nearly three times longer than the next-highest value, demonstrating its superior polaritonic properties, which again stem from its reduced linewidths at low temperatures, as in the case of the optical response.

\section{Conclusions}\label{sec3}

In this work, we systematically investigated temperature-dependent optical and polaritonic properties of high-quality, hBN-encapsulated monolayer TMDs ($\mathrm{WS_2}$, $\mathrm{MoS_2}$, $\mathrm{WSe_2}$, and $\mathrm{MoSe_2}$) from $5$ to $300\mathrm{K}$. At low temperatures, we find $\mathrm{MoSe_2}$ to exhibit notably large oscillator strengths due to its linewidth behavior. We also find that all four TMDs exhibit a negative real permittivity under these conditions, supporting surface-exciton-polaritons. Most notably, we find that $\mathrm{MoSe_2}$ exhibits the strongest polaritonic response, characterized by the largest oscillator strength, highest amplitude of negative permittivity, and optimal ratio of polaritonic response to absorption losses. Collectively, our results establish a comparative framework for understanding excitonic and polaritonic behaviors in monolayer TMDs, providing a foundation for future research aimed at exploiting the strong light-matter interactions enabled by exciton-based surface polaritons in nanoscale optoelectronic applications.

\section{Acknowledgements}
I.E. acknowledges the Israeli Science Foundation personal grant number 865/24, the Ministry of Science and Technology grant number 0005757, and the support of the European Union (ERC, TOP-BLG, Project No. 101078192). S.T acknowledges primary support from DOE-SC0020653 (excitonic testing) and NSF CBET 2330110 (environmental stabilization), additional support from Applied Materials Inc., and Lawrence Semiconductor Labs for material development / initial characterization. J.E acknowledges the support of the National Science Foundation, award number 2413808, for the hBN crystal growth.

\bibliographystyle{apsrev4-2}
\bibliography{Citings}  

\end{document}